\documentstyle[aps,preprint]{revtex}
\begin{document}

\draft

\title{Trajectory versus probability density entropy}
\author{Mauro Bologna$^{1}$, Paolo Grigolini$^{1,2,3}$, Markos
Karagiorgis$^{1}$}
\address{$^{1}$Center for Nonlinear Science, University of North Texas,\\
P.O. Box 5368, Denton, Texas 76203 }
\address{$^{2}$Istituto di Biofisica del Consiglio Nazionale delle\\
Ricerche, Via San Lorenzo 26, 56127 Pisa, Italy }
\address{$^{3}$Dipartimento di Fisica dell'Universit\`{a} di Pisa, \\
Piazza\\
Torricelli 2, 56127 Pisa, Italy }
\date{\today}
\maketitle

\begin{abstract}
We study the problem of entropy increase of the Bernoulli-shift map
without recourse to the concept of trajectory and we discuss whether,
and under
which conditions if it does, the distribution density entropy
coincides with the Kolmogorov-Sinai entropy, namely, with the
trajectory entropy.
\end{abstract}


\section{introduction}

The problem of establishing a connection between the
Kolmogorov-Sinai (KS) entropy\cite{K,S}
and the conventional entropy expressed in terms of
probability density is
an interesting problem that is attracting some attention in
literature\cite{zurek,latora}. Early work
on this subject goes back to the discussion of Goldstein and
Penrose\cite{goldstein}: These authors, almost twenty years ago,
established a connection between the KS entropy and a coarse-grained
version of the distribution density entropy.
The work of Ref.\cite{goldstein} is based on a formal and rigorous
mathematical treatment which for this reason might have eluded the
attention of
physicists working on this subject\cite{zurek,latora}. Thus we restate the
problem using intuitive arguments which also make it possible for us to
account for the more recent literature on the subject. In fact, our
heuristic treatment will allow us to relate
the results of the more recent work of Latora and
Baranger\cite{latora} to the earlier work of Zurek and Paz\cite{zurek}.

In addition to revisiting the problem of how to make the KS entropy emerge
from a nonequilibrium dynamic picture\cite{goldstein}, we shall touch also
the intriguing problem of whether a thermodynamic perspective has to
rest on the adoption of trajectories, as implied by the concept itself
of KS entropy, or on the use of
probability densities, advocated with strong arguments
by Petrosky and Prigogine\cite{tomio1,tomio2}. It is convenient
to stress that the KS entropy\cite{K,S} is a property of a single
trajectory. The phase space is divided into cells, each cell being
assigned a given label $\omega_{r}$. Then we define a sequence of
symbols by means of a single trajectory: The sequence
is determined assigning to any time step the label of the cell
where the trajectory lies at that time step. The trajectory
is supposed to be large enough
as to yield  reliable values for the
probabilities determined through the numerical frequencies. This
means that we fix a window of size $N$, and we move this window
along the sequence.
For any window position a string of symbols
$\omega_{0},\omega_{1},\ldots\omega_{N-1}$ is determined. Moving the
window of fixed size $N$ along the infinite sequence generated
 by the trajectory we have
to evaluate how many times the same string of symbols appears, thereby
leading us to determine the probability
$p(\omega_{0},\omega_{1},\ldots\omega_{N-1})$.
The KS entropy is then defined by
\begin{equation}
h_{KS} \equiv \lim_{N \rightarrow \infty} H(N)/N,
\label{definition}
\end{equation}
where $H(N)$ is the conventional Shannon entropy of the window of
size $N$ defined by
\begin{equation}
\nonumber
H(N)= \sum_{\omega_{0},,\omega_{1}\ldots\omega_{N-1}}
p(\omega_{0},\omega_{1},\ldots\omega_{N-1})
ln [p(\omega_{0},\omega_{1},\ldots\omega_{N-1})].
\label{shannonentropy}
\end{equation}
It is evident therefore that the KS entropy rests on trajectories,
and, more specifically, it implies the adoption of only one trajectory
of virtually infinite length. The KS entropy is very attractive because
its value turns out to be independent of the repartition into cells
of the phase space, due to the crucial role of the so called generating
partitions\cite{beck}. In the specific case where a natural
invariant distribution exists, it is shown\cite{pesin} that
\begin{equation}
h_{KS} = \sum_{i}\int d{\bf x}\rho_{eq}({\bf x}) \lambda_{i}({\bf x})
\label{pesintheorem}
\end{equation}
with $\lambda_{i}({\bf x}) > 0$.Note that ${\bf x}$ denotes
the coordinate of a multidimensional
phase space, $\rho_{eq}({\bf x})$ is the natural invariant
distribution and $\lambda_{i}(x)$ is a local Lyapounov coefficient,
with $i =1,d$, $d$ being the
dimension of the system under study.
From Eq.(\ref{pesintheorem}) we see that, as earlier pointed out,
the KS entropy is independent of the repartition
into cells.
The original definition of Eq.(\ref{definition}), with $N$ thought of
as time, means that the KS entropy,
as a property of a single trajectory, is the rate of
entropy increase per unit of time. However, since the single trajectory
under examination is infinitely long, and explores in time
all the phase space available, the KS entropy can also be expressed
in the form of an average over the equilibrium distribution density,
without any prejudice for the single trajectory nature of this
``thermodynamic'' property.

According to Petrosky and Prigogine\cite{tomio1,tomio2}, on the contrary,
the connection
between dynamics and thermodynamics implies the use of the Liouville
equation
\begin{equation}
\frac{\partial }{\partial t}\rho({\bf x},t) = -i L \rho({\bf x},t),
\label{liouville}
\end{equation}
where $L$ denotes both the classical and the quantum Liouville
operator, and $\rho({\bf x},t)$ is the nonequilibrium distribution density.
The reason for this choice is that the analysis of the Liouville
operator, through the ``rigged Hilbert'' space, allows the appearance of
complex eigenvalues which correspond to irreversibility, and to the
collapse of trajectories as well. This is the reason why distribution
densities are judged to be more fundamental than trajectories.

In this paper we limit our analysis to the special case where dynamics
are generated by maps rather than by Hamiltonians. We do not address
the difficult issue of discussing the thermodynamic limit $N
\rightarrow \infty$ which is the subject of very interesting recent
discussions\cite{tomio2,lebowitz}, and where, according to
Lebowitz\cite{lebowitz}, ergodicity and mixing are neither necessary
nor sufficient to guarantee the connection between dynamics and
thermodynamics. We consider the case of low-dimension chaos, where
probability emerges as a consequence of sensitivity to initial
conditions\cite{tomio2}. Even in this case, however, according to the
perspective established by Petrosky and Prigogine\cite{tomio1,tomio2},
probability densities are more fundamental than trajectories. The readers
interested in
knowing more about this perspective, entirely based on
probability density, should consult the illuminating work of
Driebe\cite{driebe}. In this case the counterpart of
Eq.(\ref{liouville}) becomes
\begin{equation}
\rho({\bf x},t+1) = \Lambda \rho({\bf x},t) ,
\label{frobenius}
\end{equation}
where $\Lambda$ is referred to as Frobenius-Perron operator. Of course,
the operator $L$ of Eq.(\ref{liouville}) has to be identified with
$i(\Lambda - 1)$.

According to the traditional wisdom, the Frobenius-Perron operator
is expected to make
the distribution densities evolve in the same way
as that resulting from the time evolution of a set of trajectories
with initial conditions determined by the
initial distribution density\cite{driebe}:The known cases of discrepancy
between the two pictures are judged to be more apparent
than real \cite{lasota}.
Nevertheless, even in the case of                             
invertible maps, the birth of irreversibility can be
studied using the same perspective as that adopted
for Hamiltonian systems, with Eq.(\ref{liouville}) replaced by
Eq.(\ref{frobenius}), and so using again probability densities
rather than trajectories.

However, we attempt at digging out the KS entropy from
Eq.(\ref{frobenius}), and this purpose forces us to formulate
a conjecture on how to
relate entropy to $\rho({\bf x},t)$. A plausible choice seems to be
\begin{equation}
S(t) = -\int_{{\bf X}}\rho({\bf x},t)ln[\rho({\bf x},t)] d{\bf x}.
\label{gibbs}
\end{equation}
We share the view of Goldstein and Penrose\cite{goldstein}
who consider the KS entropy to be a nonequilibrium entropy. In other
words, we may hope to derive the KS entropy
from the time derivative of
$S(t)$ of Eq.(\ref{gibbs}). As Goldstein and Penrose do\cite{goldstein},
to realize that purpose
we have to address a delicate problem: In the case of invertible maps,
$S(t)$ is time independent\cite{mackey}, thereby implying a vanishing
KS entropy.  Yet, the
baker's transformation, which is a well known example of invertible
map, thereby yielding a time independent $S(t)$, is shown\cite{dorfman}
to yield a KS entropy
 equal to $ln2$, a fact
suggesting
a steady condition of entropy increase. We plan to discuss all this
with the joint use of heuristic arguments and of the rigorous
theoretical tools of Ref.\cite{driebe}.

The present paper uses as a paradygm of invertible map the
two-dimensional baker's transformation, depending on two
coordinates, $x$ and $y$, the former corresponding to
dilatation and the latter to contraction. Using
this prototype for
invertible dynamics, we
aim at proving
that the adoption  of
the distribution density in the case of invertible chaotic maps would
lead to an increasing process of fragmentation, depending not only, as
the KS entropy does, on the positive Lyapounov coefficient, but also
on the negative one. The adoption of a coarse graining has the effect
of quenching the action of the negative Lyapunov coefficient, thereby
allowing the KS entropy to show up. Then, to go beyond these heuristic
arguments we make a trace on the variable $y$, namely, on the
process responsible for contraction,
and we focus our attention on the contracted dynamics. This
is equivalent to that produced by the
Bernoulli shift map. Here room is only left for dilatation and the
problem can be solved with a rigorous mathematical method, without
using trajectories.

The outline of the paper is as follows. In Section II we shall
illustrate our heuristic picture. In Section III we shall address the
problem by means of a rigorous treatment resting on the theoretical
tools provided by Driebe\cite{driebe}. In Section IV we shall draw some
conclusions. Some delicate mathematical problems behind the
theoretical calculations of Section III are detailed in Appendix.

\section{heuristic arguments}
Note that the cases studied by Latora and Baranger\cite{latora} are
two-dimensional, and our discussion here refers to a two-dimensional
case, too. We have in mind the backer's transformation and
${\bf x} \equiv (x,y)$. We denote by $W(t)$ the number of cells
occupied at a given time $t$. Note that $W(0)< W_{max}$, where the
symbol $ W_{max}$ denotes the
total number of cells into which we have divided the phase space
${\bf X}$. Our heuristic approach is based
on the following assumptions.

(i)  At
the initial time only $W(0)$ cells
are occupied.

(ii) At all times the trajectories are equally distributed
over the set of occupied cells. This
means
\begin{equation}
S(t) = ln W(t).
\label{boltzmann}
\end{equation}

(iii) We denote by $\lambda$ the positive Lyapounov coefficient, and we
set
\begin{equation}
W(t) = W(0) exp(\lambda t).
\label{plausible}
\end{equation}

All these three assumptions have been borrowed from the recent work of
Ref.\cite{zheng}. The joint use of all them yields
\begin{equation}
S(t) = \lambda t - ln W(0),
\label{KSregime}
\end{equation}
which corresponds to the Kolmogorov thermodynamical regime. Note
that the positive Lyapounov coefficient in the case of the baker's
transformation
is shown\cite{dorfman} to be:
\begin{equation}
\lambda = ln 2
\label{lyapunovequaltoln2}.
\end{equation}
Note also that according to the arguments of Section I, the connection
with the KS entropy is established through the time derivative of
$S(t)$. Thus, we conclude that
\begin{equation}
\frac{dS}{dt} = \lambda = ln 2 = h_{KS},
\label{agreement}
\end{equation}
which corresponds to deriving the KS entropy from the distribution
density picture.

This Kolmogorov
regime is not infinitely extended. It has an upper bound, given by
the fact that when equilibrium is reached, even in the merely sense
of a coarse-grained equilibrium, then the entropy stops increasing.
An estimate of this time is obviously given by the solution of the
following equation
\begin{equation}
ln W_{max} = \lambda t - ln W(0),
\label{saturation}
\end{equation}
which yields the following saturation time
\begin{equation}
t_{S}= \frac{1}{\lambda} ln (\frac{W_{max}}{W(0)})
\label{saturationtime}.
\end{equation}

Furthermore a lower bound of validity exists, which will be easily
estimated with very simple arguments. If the initial distribution
includes a large number of cells and the size of this distribution
along the coordinate $y$ is $L$, and the size of the cells is
$\epsilon$ with $\epsilon < L,$, then it is evident that, in spite of
the coarse graining the total number of cells occupied remains the
same for a while. This time is easily estimated using the equation
\begin{equation}
L exp(-\lambda t) = \epsilon,
\label{estimatebeginning}
\end{equation}
which in fact defines the time at which the distribution volume,
and consequently,the system entropy starts increasing. This time is denoted
by the symbol $t_{D}$ and reads
\begin{equation}
t_{D}= \frac{1}{\lambda} ln \left(\frac{L}{\epsilon}\right).
\label{beginningtime}
\end{equation}

We denote by $U(t)$ the volume of the distribution density at time $t$
and by $V$ the volume of the phase space, thereby implying that $U(t) \leq V$.
We note that
\begin{equation}
\frac{W_{max}}{W(0)}= \frac{V} {U(0)},
\label{volumereplacingnumber}
\end{equation}
where $V$ is the total volume of the phase space and $U(0)$ is the
initial volume of the distribution density. Thus the Kolmogorov regime
shows up in the following time interval
\begin{equation}
t_{D} = \frac{1}{\lambda} ln \left(\frac{L}{\epsilon}\right) < t <
t_{S}= \frac{1}{\lambda} ln \left(\frac{V}{U(0)}\right).
\label{validityregime}
\end{equation}
The time duration of the regime of validity of the Kolmogorov regime
can be made infinitely extended by making the cell size infinitely
small. This means that the conflict between the KS entropy
prescription and the time independence of $S(t)$ can be bypassed by
focusing our attention on the intermediate region, whose time duration
tends to infinity with $\epsilon \rightarrow 0 $. We note that a
choice can be made such that $V/U(0) = (L/\epsilon)^{\chi}$, with
$\chi >1$. This means the time duration of the Kolmogorov regime
can be made $\chi$ times larger than the time duration
of the transition regime. For $\epsilon \rightarrow 0$ both time
durations become infinite, thereby showing that a Kolmogorov regime
of infinite time duration can be obtained at the price, however, of
waiting an infinitely long time for the entropy to increase.
The infinite waiting time before the regime of entropy increase fits
the observation\cite{mackey,lasota} that the Gibbs entropy of an
invertible map is constant. The
linear entropy increase showing up ``after this infinite waiting time''
allows the emergence of the KS entropy from within the probability
density perspective.

This kind of coarse graining might be criticized as corresponding to
arbitrary choices of the observer.
 It is interesting to remark that there exists another interesting
form of coarse graining, produced by weak stochastic forces. Both in
the case where this stochastic forces mimic the interaction with the
environment\cite{zurek} or in the case where it happens to be an
expression of spontaneous fluctuations\cite{luca} this kind of coarse
graining can be regarded as being produced by nature.
 Here we limit ourselves to remarking that according
to Zurek and Paz\cite{zurek} these stochastic forces contribute  a
fluctuation-dissipation process mimicking the interaction between the
system of interest and the environment. These authors studied the
inverted stochastic oscillator
\begin{equation}\label{paolo18}
\frac{d^{2}x}{dt^{2}} = \lambda^{2} x(t) + \gamma \frac{dx}{dt} + f(t),
\label{inverted}
\end{equation}
where the friction $\gamma$ and the stochastic force $f(t)$ are
related to one another by the standard fluctuation-dissipation
relation
\begin{equation}
<f f(t)> = 2 \gamma <(\frac{dx}{dt})^{2}>_{eq} \delta(t) \equiv 2 D
\delta(t).
\label{standard}
\end{equation}
It is interesting to remark that the proper formulation of the second
principle implies that the entropy of a system can only increase or remain
constant under the condition of no energy exchange between the system
and its environment. In the case of Eq.(\ref{standard}) the
energy exchange between system and environment is negligible for any
observation made in the time scale
\begin{equation}\label{tempo}
t << 1/\gamma
\label{noenergyexchange}.
\end{equation}
To ensure that the system entropy increase to
take place with no energy exchange between system and its environment Zurek
and Paz [3] set the condition of Eq.(\ref{tempo}) and this, in turn, allows them to
neglect the friction term in Eq.(\ref{paolo18}).
 Then, these authors adopted the modes
\begin{equation}
u \equiv \frac{dx}{dt} + \lambda x
\label{firstmode}
\end{equation}
and
\begin{equation}
w \equiv \frac{dx}{dt} + \lambda x,
\label{secondmode}
\end{equation}
which make it possible for them to split Eq.(\ref{inverted}) into
\begin{equation}
\frac{du}{dt} = \lambda u(t) + f(t)
\label{splitting1}
\end{equation}
and
\begin{equation}
\frac{dw}{dt} = -\lambda w(t) + f(t)
\label{splitting2}.
\end{equation}

Let us imagine the initial distribution density as a rectangle of
size $\Delta w(0)$ along the direction $w$ and $\Delta u(0)$ along the
direction
$u$. We keep denoting by $U(t)$ the distribution
volume at a given time $t$. Thus the
volume of the initial distribution
is
 \begin{equation}
 U(0)  = \Delta u(0)\Delta w(0).
 \label{initialvolume}
 \end{equation}
In the absence of the stochastic force
$f(t)$, Eqs.(\ref{splitting1}) and Eqs.(\ref{splitting1}) result in an
exponential increase and an exponential decrease, with the same rate
$\lambda$, respectively. Consequently, the Liouville theorem $U(t) =
U(0)$ is fulfiled.
In the presence of stochastic force, we work as follows. In the former
equation,
with $u$ increasing beyond any limit,
 the weak stochastic force $f(t)$ can be neglected. This is not the
 case with the latter equation. In fact, $w$ is a contracting variable
 in the absence of the stochastic force. In the presence of the
 stochastic force the minimum size of the distribution along $w$ is
 given by
 \begin{equation}
 <w^{2}>_{eq}^{1/2} = (D/\lambda)^{1/2}.
 \label{minimumsize}
 \end{equation}
 This minimum size is reached in a time determined by the solution of
 the following equation
 \begin{equation}
 \Delta w(0)exp(-\lambda t) = (D/\lambda)^{1/2}
 \label{balance}
 \end{equation}
 yielding
 \begin{equation}
 t_{D} = \frac{1}{\lambda} ln \left(\frac{\lambda}{D}\right)^{1/2} \Delta w(0).
 \label{transitiontimewithstochasticforce}
 \end{equation}

 Due to the fact that deterministic chaos is simulated by Zurek and
 Paz\cite{zurek} by means of an inverted parabola, these authors did
 not consider the entropy saturation effects. However, it is
 straigthforward to evaluate the saturation effect with heuristic
 arguments concerning the case where the total volume of the phase
 space has the finite value $V$. From the time $t = t_{D}$ on, the
 distribution volume $U(t)$ increases exponentially in time with the
 following expression
 \begin{equation}
 U(t) = \Delta w(0) \Delta u (0) exp(\lambda t)
 = (D/\lambda)^{{1/2}} \Delta u(t_{D}) exp(\lambda t).
 \label{increase}
 \end{equation}
 Thus, the saturation time is now given by
 \begin{equation}
 t_{S} = \frac{1}{\lambda} ln \left[\frac{V}{\Delta u(0) \Delta
 w(0)}\right].
 \label{saturationtime2}
 \end{equation}
 Using Eq.(\ref{initialvolume}) we can write
 this saturation time as
 \begin{equation}
 t_{S} =  \frac{1}{\lambda} ln \left[\frac{V}{U(0)}\right],
 \label{alternative}
 \end{equation}
 which coincides with Eqs.(\ref{validityregime}) and
 (\ref{saturationtime}).

  In conclusion, it seems that the emergence of a Kolmogorov regime is
 made possible by the existence of a form of coarse graining, and
 that it is independent of whether the coarse graining is realized by
 the division into cells or by a weak stochastic force. This property
 seems to make less important the discussion of whether the stochastic
 force is of environmental origin or rests on some kind of extension
 of the current physical laws. However, we have to point out that the
 situation significantly changes, if we move from a strongly to a
 weakly chaotic classical system. As a relevant example, let us refer
 ourselves to the work of Ref.\cite{elena}. The authors of this work
 study the asymptotic time limit of a diffusion process generated by using an
intermittent map as a
dynamic generator of diffusion. If
 these dynamics are perturbed by a white noise, a transition
 is provoked,
  at long
 times, from anomalous to normal diffusion. When the only source of
 random behavior is given by the sporadic randomness of the
 intermittent map\cite{bologna}, the long-time limit is characterized
 by L\'{e}vy statistics, a physical condition in a striking
 conflict with the condition of Gaussian statistics produced by the
 action of fluctuations\cite{elena}.
Here we limit our attention to the case of strong chaos where the two
distinct sources of coarse graining produce equivalent effects. It
might be of some interest for the reader to compare the coarse-graining
approach of this section to the more formal method recently
adopted by Fox\cite{fox} to deal with the same problem.

It is interesting to stress that to make the regime of validity of
the Kolmogorov regime as extended as possible we must make the
ratio $V/U(0)$ as large as possible (virtually infinite). This means
that we have to choose an initial distribution density so sharp as to
become apparently equivalent to a single trajectory. This seems to
be an attractive way of explaining why in this condition the KS
entropy is recovered, since, as stressed in Section I, the KS entropy
is a single trajectory property. However, in accordance with the
authors of Refs.\cite{tomio1,tomio2,driebe} we must admit that there
exists a deep difference between a trajectory and a very sharp
distribution. The latter is stable and robust, while the former is not.
In Section III we shall show that the rigorous derivation of the
Kolmogorov regime requires a non trivial mathematical procedure, and
the mathematical effort to make from this side, to derive the KS
entropy, serves the useful purpose
of proving  that the KS entropy of a trajectory is a really wise way of
converting into advantages the  drawbacks of the trajectory
instability.

\section{The KS entropy from a reduced Frobenius-Perron equation}

This Section is devoted to a rigorous discussion
resting only on the theoretical tools
described in Ref.\cite{driebe} for a
genuine probability density aproach.
According to Mackey\cite{mackey}, if we rule out the
possibility that the laws of physics
are misrepresented by invertible dynamic prescriptions, there are only
two possible sources of entropy increase. The first is the coarse
graining discussed in Section II. The second is the adoption of
reduced equation of motion, obtained by a trace over ``irrelevant''
degrees of freedom\cite{mackey}. In fact here we
study the Bernoulli shift map,
\begin{equation}
x_{t+1} =2 x_{t}  ,    mod 1 .
\label{Bernoulli}
\end{equation}
The Frobenius-Perron equation of this map is defined by\cite{driebe}
\begin{equation}
\rho(x,t+1) = \Lambda \rho(x,t) \equiv
\frac{1}{2} [\rho(\frac{x}{2},t) + \rho(\frac{x+1}{2},t)].
\label{end2}
\end{equation}
It is straigtforward to show that the Frobenius-Perron operator
of Eq.(\ref{end2}) stems from the contraction over the variable $y$ of the
 baker's mapping, acting in fact on the unit square of
two-dimensional space (x,y) (see, for instance Ref.\cite{dorfman}).
It is shown\cite{dorfman} that the KS entropy of the baker's
transformation is well defined and turns out to be the same as that
of the Bernoulli shift map, namely $h_{KS}= ln2$. Intuitively, this
suggests that the main role of coarse graining is that of making
inactive the
process of contraction, and with it the
negative Lyapunov coeficient. This intuitive argument seems
to be plausible and raises the interesting
question of how to prove it with a rigorous approach. This
is equivalent to deriving the Kolmogorov regime using a rigorous
mathematical method rather than the heuristic arguments of
Section II. We must observe again that this is made possible by the
fact that the tracing has changed the originally invertible map into
one that is not invertible.

To address this issue we follow the prescription of Ref.\cite{driebe}. 
First of all, we express the distribution density at time $t$ under the form given by Ref. \cite{driebe} 
which reads:
\begin{equation}
\rho(x,t) = 1+\sum_{j=1}^{\infty}exp(-\gamma_{j}t)\frac{B_{j}(x)}{j!}
[\rho^{(j-1)}(1,0) - \rho^{(j-1)}(0,0)].
\label{mauro1}
\end{equation}
Note that $\gamma_{j} \equiv j ln2$, $B_{j}(x)$ are the Bernoulli
polynomials\cite{abramowitz} and $\rho^{(n)}(x,t)$ denotes the $n$-th order
derivative of $\rho(x,t)$ with respect to $x$. Hereby, we shall show how to derive
from the previous one more tractable expression,which will be checked in
appendix.

 In the case of an initial condition close to equilibrium,
resulting from the sum of the equilibrium distribution and the first
``excited'' state, it is easy to prove that the entropy $S(t)$ of
Eq. (\ref{gibbs}) reaches exponentially in time the steady-state
condition. This  suggests that the Kolmogorov regime, where the
entropy $S(t)$ is expected to be a linear function of time,  must
imply an initial condition with infinitely many ``excited''
states.
To deal with a condition of this kind it is convenient to express
Eq.(\ref{mauro1}) in an equivalent form given by
\begin{equation}
\rho(x,t) =  1 + \sum_{j=1}^{\infty}
\int_{-\infty}^{+\infty}\frac{B_{j}(x)}{j!}(-iz\omega)^{{j-1}}\hat{\rho}(\omega)
(exp(-i\omega) - 1)\frac{d\omega}{2\pi},
\label{mauro2}
\end{equation}
where $z \equiv exp[- t(ln2)]$ and $\hat{\rho}(\omega)$ is related
to the initial condition
$\rho(x,0)$ by the Fourier transform
\begin{equation}
\rho(x,0) =
\int_{-\infty}^{+\infty}\hat{\rho}(\omega)exp(-i\omega x)
\frac{d\omega}{2\pi}.
\label{mauro3}
\end{equation}
The following equation
\begin{equation}
\sum_{j = 0}^{\infty} \frac{B_{j}(x)}{j!} z^{j} =
z\frac{exp(zx)}{exp(z)-1},
\label{mauro4}
\end{equation}
is known\cite{abramowitz} to generate Bernoulli polynomials. Using
this Bernoulli polynomial generatrix,
we arrive, after some algebra, at
\begin{equation}
\rho(x,t) = z \int_{-\infty}^{+\infty}exp(-i\omega zx)\hat{\rho}(\omega)
\frac{exp(-i\omega) -1}{exp(-i\omega z) -1} \frac{d\omega}{2\pi}.
\label{mauro5}
\end{equation}
By expanding the denominator of Eq.(\ref{mauro5}) into a Taylor series
and using Eq.(\ref{mauro3}), we finally derive the fundamental
expression
\begin{equation}
\rho(x,t) = z \sum_{n=0}^{\infty}[\rho(zx+zn, 0) -\rho(zx+zn+1, 0)].
\label{mauro6}
\end{equation}
This important expression makes it possible for us to discuss
analytically the entropy time evolution ensuing the preparation of an
initially very sharp distribution. Let us consider in fact
\begin{equation}
\rho(x,0) = \frac{\alpha}{1-exp(-\alpha)} exp(-\alpha x), 0 \leq x
\leq 1.
\label{mauro7}
\end{equation}
For $\alpha \rightarrow \infty$ this initial distribution becomes a
very sharp distribution located at $x=0$. By plugging this initial
distribution density into Eq.(\ref{mauro6}) we obtain
\begin{equation}
\rho(x,t) = z \alpha \frac{exp(-\alpha x z)}{1- exp(-\alpha  z)}.
\label{mauro8}
\end{equation}
It is evident that this simple analytical expression for the time
evolution of the distribution density is exact, and corresponds to the
time evolution dictated by the Frobenius-Perron operator of
Eq.(\ref{end2}).

We are now in a position to discuss the central issue of this paper,
namely, the time evolution of the Gibbs entropy of Eq.(\ref{gibbs}),
which, in the case here under study, reads
\begin{equation}
S(t) = -\int_{X}\rho(x,t)ln[\rho(x,t)] dx,
\label{mauro9}
\end{equation}
with $X$ now denoting  the interval $[0,1]$.
By plugging Eq.(\ref{mauro7}) within Eq.(\ref{mauro9}) we obtain
\begin{equation}
S(t) = 1 - ln(\alpha z)+ \ln \left[ 1-exp(- \alpha z) \right]- \frac{\alpha
z}{exp(\alpha z) -1} .
\label{mauro10}
\end{equation}
In the limiting case $\alpha\rightarrow \infty$ this exact prediction
is approximated very well by
\begin{equation}
S(t) = -ln(\alpha) + (ln2) t.
\label{mauro11}
\end{equation}
It indicates that a sharp
initial distribution makes the system evolve according to the KS
entropy, with no  regime of transition from mechanics to
thermodynamics. The third regime of
Ref.\cite{latora} is still present. It is straigtforward to show that
the saturation time $t_{S} = ln{\alpha}/ln2$ resulting from
Eq.(\ref{mauro10}) is the same as that of Eq.(\ref{saturationtime}) in the case
$V=1$. In fact using Eq.(\ref{volumereplacingnumber}) and $V = 1$
we obtain that $W_{max}/W(0) = 1/U(0)$, where $U(0)$ is the size of the
initial distribution. The size of the initial distribution of
Eq.(\ref{mauro7}), for $\alpha \rightarrow \infty$, becomes
prportional to $1/\alpha$. Thus, $ln\alpha \approx ln (W_{max}/W(0)$
in accordance with Eq. (\ref{saturationtime}).

This is an elegant result, involving a modest amount of algebra.
However, it refers to an initial distribution located at $x=0$.
We want to prove that this is a general property, independent of where
the initially sharp distribution is located, at the price, as we shall
see, of a more complicated mathematical treatment.
For this purpose we study the case where the distribution shape
is the Lorentzian curve:
\begin{equation}\label{lorent}
\rho\left(x,0\right) =A
\frac{\Gamma}{\left(x-x_{0}\right)^{2}+\Gamma^{2}},
\end{equation}
with $ x_{0}$ being a generic point of the
interval $ \left[0,1 \right]$ and $x$ running in the same interval.
Setting the normalization condition yields
\begin{equation}\label{normal}
A=\frac{1}{\arctan \left(\frac{x_{0}}{\Gamma }
\right)+\arctan \left(\frac{1-x_{0}}{\Gamma } \right)}.
\end{equation}
We have to set again the condition that the initial distribution is
very sharp. Thus we make the assumption $ \Gamma\rightarrow 0 $,
yielding  $ A \approx 1/\pi  $.
We plug this approximated value of $A$ into  Eq.  (\ref{mauro6}),
thereby obtaining the following density time evolution
\begin{equation}\label{rhot}
\rho \left(x,t \right) = \frac{z\Gamma }{\pi}
\sum _{n=0}^{\infty }\left[\frac{1 }{\left(zx+zn-x_{0}\right)^{2}+\Gamma^{2}} -
\frac{1 }{\left(zx+zn-x_{0}+1\right)^{2}+\Gamma^{2}}\right].
\end{equation}

We are now in a position to study the entropy time evolution again.
Plugging Eq.(\ref{rhot}) into  (\ref{mauro9}) we find
\begin{eqnarray*}
S\left(t \right) & =&  - \int\limits_{X}\rho \left(x,t
\right)\ln\left[\frac{z}{\pi \Gamma }
\sum _{n=0}^{\infty }\frac{1 }{\left(\frac{zx+zn-x_{0}}{\Gamma}
\right)^{2}+1} -
\frac{1 }{\left(\frac{zx+zn-x_{0}+1}{\Gamma} \right)^{2}+1}\right]dx
\\ = -  \int\limits_{X}\rho \left(x,t \right)\ln\left[\frac{z}{\pi \Gamma
}\right]dx   &  -     &
 \int\limits_{X}\rho \left(x,t \right)\ln\left[
\sum _{n=0}^{\infty }\frac{1 }{\left(\frac{zx+zn-x_{0}}{\Gamma}
\right)^{2}+1} -
\frac{1 }{\left(\frac{zx+zn-x_{0}+1}{\Gamma} \right)^{2}+1}\right]dx ,
\\
\end{eqnarray*}
where $[y]$ denotes the integer part of $y$.
To derive a more tractable expression
 we note that  in the limiting case  of $\Gamma$  very small, the
 quantities $[\ldots ]$
 contributing  this series
are almost zero except for
$ n=-\left[x \right] +\left[\frac{x_{0}}{z} \right]
=  \left[\frac{x_{0}}{z} \right]$  in the first term, and
for the possible contribution
$ n=-\left[x \right] -\left[\frac{1-x_{0}}{z} \right]
=  -\left[\frac{1-x_{0}}{z} \right]$ in the second term.
The latter condition cannot be realized, since $n$ is a positive integer.
Thus, using only the first class of  contributions,
we get for the entropy $S(t)$ the following approximate expression
\begin{equation}\label{aprentrloren}
S\left(t \right)\approx \ln\Gamma +\left(\ln2 \right)t -\ln\pi-
\int\limits_{0}^{z/\Gamma}\frac{\ln\left(y^{2}+1 \right)}{y^{2}+1}dy,
\end{equation}
which, in the limiting case $ \frac{z}{\Gamma }\rightarrow \infty $
becomes
\begin{equation}\label{aprfinal}
S\left(t \right)\approx \ln\Gamma +\left(\ln2 \right)t -\ln\pi-
\int\limits_{0}^{\infty}\frac{\ln\left(y^{2}+1 \right)}{y^{2}+1}dy
= \ln\Gamma +\left(\ln2 \right)t-\ln\pi-\pi\ln2\approx
\ln\Gamma +\left(\ln2 \right)t.
\end{equation}

As in the earlier case, the validity of the approximation yielding
the linear dependence of $S(t)$ on time,
is broken
at the time $ t\sim -\ln\Gamma /\ln2$.
In conclusion,
the Kolmogorov condition is realized
by very sharp initial distributions.
Our derivation of this interesting property was done adhering to the
recommendation of Ref.\cite{tomio1,tomio2,driebe} of resting
only on densities rather
than on trajectories.

\section{concluding remarks}

This paper shows that there exists a subtle difference between a
Liouville density and a probability distribution. The Liouville
distribution density of a chaotic map, which is at the same time
invertible,
 becomes increasingly
fragmented with time. If the initial distribution has a volume which
is much smaller than the volume of the phase space, the highly
fragmented distribution density at large times has a volume identical
to the initial, but the impression
afforded by a coarse-grained observation is that the
initial volume increases with time till to become equal to that of the
whole phase space. If we do the calculation of the time evolution of the
Liouville density observing the motion of many trajectories
with slightly different initial conditions, we are forced to adopt a
coarse-graining procedure, dictated by the need itself of counting
how many trajectories are found at a given time in a given small
region of the phase space. This has the effect of making the Gibbs
entropy increase. If, on the contrary, the calculation genuinely rests
on the motion of the Liouville density, thereby implying that a
quantum-like formalism is adopted, the Gibbs entropy is constant.

On the other hand, the KS entropy is a trajectory property and this,
to first sight, might lead us to believe that, being a trajectory
property, cannot be recovered from within
an approach genuinely resting on the distribution density.
The heuristic arguments used in Section II subtly rest on the
trajectory properties and thus on the assumption that
the two perspectives are equivalent in spite of the warnings of the
authors of Refs.\cite{tomio1,tomio2,driebe}. This is the reason why
we judge the  theoretical calculations of Section III to be
of significant interest. These results have been obtained
without having any direct recourse to the trajectory instability
and only using the theoretical tools illustrated in Ref.\cite{driebe},
which in fact address the problem of the dynamics of map using the
quantum mechanical language of eigenstate and eigenvalue.
It has to be pointed out that to establish this rigorous connection
between the linear increase of the Gibbs entropy and the KS entropy
we need to use infinitely many ``excited'' states, namely, a condition
very far from equilibrium. The adoption of a ``reduced'' Liouville-like
equation has been essential for the success of this enterprise.
The Frobenius-Perron operator of
the Bernoulli shift map is obtained from the Frobenius-Perron operator
of the baker's transformation via contraction over the variable $y$,
corresponding to the contraction process. As a consequence, room is
only left for dilatation. This is the reason why the heuristic
argument of Eq.(\ref{plausible}) holds true with no restriction and
the use of a method rigorously based on densities lead to the same
result. After all, the adoption of the theoretical tools of
Ref.\cite{driebe} as a rigorous way to evaluate the regression to
equilibrium, must reflect in a way the trajectory instabilities behind  the
Pesin
theorem of Eq.(\ref{pesintheorem}).

In the case where the action of the negative Lyapounov coefficient is
quenched by a coarse-graining process, in addition to the saturation
and to the Kolmogorov regime,
also a short-time regime of transition to thermodynamics
appears, so that, as found in Ref.\cite{latora} three distinct
regimes can be detected, the regime of transition to thermodynamics,
the Kolmogorov regime, and the saturation regime. Here we show that
the three regimes discussed by Latora and Baranger\cite{latora} are
exhibited also in the case of a coarse graining produced by random
fluctuations.

In conclusion, this paper contributes to deepening our understanding
of the connection between density and trajectory distribution. The
density entropy exhibits a regime of increase linear in time
with a rate equivalent to that of the trajectory entropy if the
initial distribution is very sharp. It has to stressed that it cannot
be infinitely sharp. This would make the distribution density
useless, since an infinitely sharp
initial distribution would manifest the lack of robustness pointed
out by Driebe\cite{driebe}. Thus, the connection is established
using a genuine density. At the same time
the accordance between the heuristic approach yielding
the  derivation of Eq.(\ref{mauro11})
can be interpreted as a rigorous support of the assumptions
made by some some authors\cite{zheng} to derive the KS entropy
from within a probability distribution approach.

We want to remark that we are not aware of earlier work where the
Lyapounov coefficient is derived analytically from a probability
density approach with no use of heuristic arguments but that of
Pattanayak and Brumer\cite{pattanayak} which, however, seems to be
limited to the study of the transition regime\cite{brumer}.

A problem open to future research concerns the role of coarse-graining.
We have seen that in the case of strong chaos studied in this paper
there is no essential difference between the coarse graining resulting
from the repartition of the phase space into cells and the coarse
graining caused by fluctuations. In the case of dynamic systems like
the weakly chaotic billiards studied by
Zaslavsky\cite{zaslavsky,today}, which are easily proved to be
statistically equivalent to the intermittent maps of Ref.\cite{elena},
the two different coarse-graining sources result in different physical
effects at long times. On the other hand, it is expected\cite{massi}
that the dynamic process of transition to L\'{e}vy statistics is
characterized by thermodynamic properties of non-extensive
nature\cite{tsallis}. The study of the time evolution of
non-extensive entropy under
the actions of the two different sources of coarse graining studied in
this paper
would an interesting program for future research
work.${}\\$

\setcounter{equation}{0}
\renewcommand{\theequation}{A-\arabic{equation}}
\appendix{\textbf{APPENDIX}}
${}\\$

The purpose of this Appendix is to check the main  result of
Section III. Our theoretical reference on this issues is given by the book of Ref. \cite{driebe}. We
note, however, that 
Eqs. (\ref{mauro8}) and Eq.(\ref{rhot}) are not directly derived from Eq.(\ref{mauro1}), which is a theoretical finding of 
Ref. \cite{driebe}, but they are derived from Eq.(\ref{mauro6}), wich is the result of a further development of the theory of 
Ref. \cite{driebe}. We feel therefore the need of proving that Eq.(\ref{mauro6}) fits the main requirement of keeping the norm unchanged and of being an exact solution of Frobenius-Perron equation of Eq.(\ref{end2}). To double check our results, we shall prove also that Eqs. (\ref{mauro8}) and Eq.(\ref{rhot}) fit the same property. As
a general remark about the content of this appendix we note
that a function of the variable $x$, defined only within the
finite interval $[0,1]$, admits a treatment based on its Fourier
transform if it is thought of as being defined on the whole
interval $[-\infty,\infty]$ with vanishing value outside
 $[0,1]$. Similarly we can define the Fourier series of this function
assuming it to be periodically repeated all over the real axis. We
shall adopt this approach throughout the whole appendix.\par
    Let us check Eq.(\ref{mauro6}) first. We note that the argument of the
density of Eq.(\ref{mauro6}) can be arbitrary with the only
condition that the variable $x$ is in the interval [0,1].
Furthermore, since Eq.(\ref{mauro6}) is derived from
Eq.(\ref{mauro5}), it is enough for us to prove that
Eq.(\ref{mauro5}) is properly normalized and is a solution of the
Frobenius-Perron equation of Eq.(\ref{end2}). In conclusion, we
have to check:
\begin{equation} \label{mau5}
\rho(x,t) = z \int_{-\infty}^{+\infty}e^{-i\omega
zx}\hat{\rho}(\omega) \frac{e^{-i\omega} -1}{e^{-i\omega z} -1}
\frac{d\omega}{2\pi}.
\end{equation}
First we check that this equation is norm conserving or:
\begin{equation}
\int^{1}_{0}\rho\left(x,t\right)dx=1.
\end{equation}
To do so, we integrate Eq.(\ref{mau5} with respect to the variable
x from $-\infty$ to $+\infty$. Thus we obtain:
 \begin{eqnarray} \label{mauro}
\int^{1}_{0} z \int_{-\infty}^{+\infty}e^{-i\omega
zx}\hat{\rho}(\omega) \frac{e^{-i\omega} -1}{e^{-i\omega z} -1}
\frac{d\omega}{2\pi}&dx&
\nonumber
\\
=\int_{-\infty}^{+\infty}\frac{e^{-i\omega
z}-1}{-\imath\omega}\hat{\rho}(\omega) \frac{e^{-i\omega}-1}{e^{-i\omega z} -1} \frac{d\omega}{2\pi}.
 \end{eqnarray}
This means the expression
\begin{equation}\label{contr1}
\int_{-\infty}^{+\infty}\frac{e^{-i\omega}-1}{-\imath\omega}\hat{\rho}(\omega)
\frac{d\omega}{2\pi}\equiv\int^{1}_{0}\int_{-\infty}^{+\infty}e^{-i\omega y }\hat{\rho}(\omega)\frac{d\omega}{2\pi}dy.
\end{equation}
The integral over $\omega$ is by definition the Fourier transform
of $\rho(y,0)$, yielding thereby
\begin{eqnarray} \label{contr2}
\nonumber
\\
\int^{1}_{0}\rho\left(x,t\right)dx=\int^{1}_{0}\rho(y,0)dy=1\,\forall
t,
\end{eqnarray}
due to the fact that the initial condition is assumed to be
normalized. \par
 We want now to prove that the distribution
density $\rho(x,t)$ of Eq. (\ref{mau5}) is a solution of the
Frobenius-Peron operator of Eq.(\ref{end2}), namely, that:

\begin{eqnarray} \label{contr3}
\rho(x,t+1)=\frac{1}{2}\left[\rho(\frac{x}{2},t)+\rho(\frac{x+1}{2},t)\right].
\nonumber
\\
\end{eqnarray}
Remembering that $z=2^{-t}$ and  $\frac{z}{2}=2^{-t-1}$ we can write
Eq.(\ref{mau5})as:
\begin{eqnarray} \label{contr4}
\rho(x,t+1) = \frac{z}{2} \int_{-\infty}^{+\infty}e^{-i\omega zx/2}\hat{\rho}(\omega) \frac{e^{-i\omega} -1}{e^{-i\omega z/2}
-1} \frac{d\omega}{2\pi}.
\end{eqnarray}
Plugging Eq.(\ref{mau5}) into it, the r.h.s. of Eq.(\ref{contr3})
becomes:

\begin{displaymath} \label{contr5bis}
\frac{1}{2}\left[\rho(\frac{x}{2},t)+\rho(\frac{x+1}{2},t)\right]
\end{displaymath}
\begin{eqnarray} \label{contr5}
=\frac{z}{2}\left[ \int_{-\infty}^{+\infty}e^{-i\omega
zx/2}\hat{\rho}(\omega) \frac{e^{-i\omega } -1}{e^{-i\omega z} -1}
\frac{d\omega}{2\pi}+\int_{-\infty}^{+\infty}e^{-i\omega
z(x+1)/2}\hat{\rho}(\omega) \frac{e^{-i\omega } -1}{e^{-i\omega z}
-1} \frac{d\omega}{2\pi}\right]&
\end{eqnarray}
and after a little algebra we get:
\begin{equation} \label{contr6}
\frac{1}{2}\left[\rho(\frac{x}{2},t)+\rho(\frac{x+1}{2},t)\right]=
\frac{z}{2} \int_{-\infty}^{+\infty}e^{-i\omega zx/2}
\hat{\rho}(\omega) \frac{\left[e^{-i\omega }
-1\right]\left[e^{-i\omega zx/2}+1\right]}{e^{-i\omega z} -1}
\frac{d\omega}{2\pi}.
\end{equation}
By decomposing the denominator as follows
\begin{equation} \label{denom}
\frac{1}{e^{-i\omega z} -1}= \frac{1}{\left[e^{-i\omega z/2}
-1\right]\left[e^{-i\omega z/2} +1\right]}
\end{equation}
and simplifying, we obtain
\begin{equation} \label{contr7}
\frac{1}{2}\left[\rho(\frac{x}{2},t)+\rho(\frac{x+1}{2},t)\right]=
\frac{z}{2} \int_{-\infty}^{+\infty}e^{-i\omega zx/2}
\hat{\rho}(\omega) \frac{e^{-i\omega } -1}{e^{-i\omega z/2}
-1} \frac{d\omega}{2\pi}
\end{equation}
that coincide with Eq.(\ref{contr4}).

To check the property $ \rho \left(x,t \right)\to 1$ for $ t\to \infty $ or
$ z\to 0$ we use (\ref{mau5}) and we write:
\begin{eqnarray}\label{asyntro}
\nonumber
\rho \left(x,t \right)\sim z \int_{-\infty}^{+\infty}\hat{\rho}(\omega) \frac{e^{-i\omega} -1}{-i\omega z }
\frac{d\omega}{2\pi}&
\\
\nonumber
 =\int_{-\infty}^{+\infty}\hat{\rho}
 (\omega) \frac{e^{-i\omega} -1}{-i\omega }\frac{d\omega}{2\pi}=
 \int_{-\infty}^{+\infty}\hat{\rho}
 (\omega) \int_{0}^{1} e^{-i\omega x}dx\frac{d\omega}{2\pi}=&
\int_{0}^{1}\rho(x) dx=1.
\end{eqnarray}
\par
Now we shall test directly the  Eq.(\ref{mauro8}) using
Eq.(\ref{mauro1}). Plugging directly the initial distribution:
\begin{equation}
\rho(x,0) = \frac{\alpha}{1-e^{-\alpha}} e^{-\alpha x}, 0 \leq x
\leq 1
\end{equation}
into Eq.(\ref{mauro1}) we obtain:
\begin{eqnarray}
\rho(x,t) = \sum_{j=0}^{\infty}e^{-\gamma_{j}t}\frac{B_{j}(x)}{j!}
[\rho^{(j-1)}(1,0) - \rho^{(j-1)}(0,0)]\nonumber
\\
=\frac{\alpha}{1-e^{-\alpha}}
\sum_{j=0}^{\infty}e^{-\gamma_{j}t}\frac{B_{j}(x)}{j!}
[(-\alpha^{j-1})e^{-\alpha} -(-\alpha^{j-1})].
\end{eqnarray}
Using the Bernoulli polynomials generatrix \cite{abramowitz} we
get
\begin{equation}
\rho(x,t) = z \alpha \frac{e^{-\alpha x z}}{1- e^{-\alpha  z}}.
\label{mau8}
\end{equation}
that coincides with Eq.(\ref{mauro8}) obtaned using the formula Eq.(\ref{mauro6})

Finally let us to check the norm conservation of Eq.(\ref{rhot}).
Without any
approximation, using the value of $A$ of
Eq.(\ref{normal}), and a little algebra, we get:
\begin{eqnarray}
&\int_{0 }^{1}\rho\left(x,t\right)\,dx =\left[\arctan\left( \frac{x_{0}}{\Gamma}\right)+
\arctan\left(\frac{1- x_{0}}{\Gamma}\right)\right]^{-1}
\nonumber
\\
&\cdot
\int_{0 }^{1}
z\Gamma\sum _{n=0}^{\infty } \left[\frac{1
}{\left(zx+zn-x_{0}\right)^{2}+\Gamma^{2}} -\frac{1
}{\left(zx+zn-x_{0}+1\right)^{2}+\Gamma^{2}}\right]dx 
\nonumber
\\
&=\left[\arctan\left( \frac{x_{0}}{\Gamma}\right)+
\arctan\left(\frac{1- x_{0}}{\Gamma}\right)\right]^{-1}\cdot
\sum^{+\infty}_{n=0}\left[
\arctan\left(\frac{xz+nz-x_{0}}{\Gamma}\right)-
\arctan\left(\frac{xz+nz+1-x_{0}}{\Gamma}\right)\right]^{x=1}_{x=0}.
\end{eqnarray}
This yields:
\begin{eqnarray*}
&\int_{0 }^{1}\rho\left(x,t\right)\,dx =\left[\arctan\left( \frac{x_{0}}{\Gamma}\right)+
\arctan\left(\frac{1- x_{0}}{\Gamma}\right)\right]^{-1}
\nonumber
\\
&\cdot\sum^{+\infty}_{n=0}\biggl[
\arctan\left(\frac{z+nz-x_{0}}{\Gamma}\right)-
\arctan\left(\frac{z+nz+1-x_{0}}{\Gamma}\right)-\arctan\left(\frac{nz-x_{0}}{\Gamma}\right)+
\arctan\left(\frac{nz+1-x_{0}}{\Gamma}\right)\biggr].
\end{eqnarray*}
Examining the expression we can note that only the terms with
$n=0$  survive in the sum but that terms simplify with the external factor (the constant
$A$) so finally
\begin{equation}
\int_{0 }^{1}\rho\left(x,t\right)\,dx =1\, \forall t.
\end{equation}
Checking that Eq.(\ref{rhot}) fulfils the Frobenius-Perron
operator involves some extended but straightoforward algebra.

\end{document}